\documentclass[journal,twoside,web]{ieeecolor}
\usepackage{generic}
\usepackage{cite}
\usepackage{amsmath,amssymb,amsfonts}
\usepackage{algorithmic}
\usepackage{graphicx}
\usepackage{textcomp}
\usepackage{pifont}

\newcommand{\xmark}{\ding{55}}
\usepackage{framed,multirow, multicol}
\usepackage{latexsym}
\usepackage{url}
\usepackage{xcolor}
\usepackage{hyperref}
\usepackage{booktabs}
\usepackage{mathtools}
\usepackage{caption}

\def\BibTeX{{\rm B\kern-.05em{\sc i\kern-.025em b}\kern-.08em
    T\kern-.1667em\lower.7ex\hbox{E}\kern-.125emX}}

\begin{document}
\title{Generation of Structurally Realistic Retinal Fundus Images with Diffusion Models}

\author{Sojung Go, Younghoon Ji, Sang Jun Park, Soochahn Lee 
\thanks{Corresponding authors: Sang Jun Park, Soochahn Lee.}
\thanks{Sojung Go, Sang Jun Park are with the Department of Ophthalmology, Seoul National University College of Medicine, Seoul National University Bundang Hospital, Seongnam, Korea.}
\thanks{Younghoon Ji is with the VUNO, Inc., Seoul, Korea.}
\thanks{Soochahn Lee is with the School of Electrical Engineering, Kookmin University, Seoul, Korea.}}

\maketitle

\begin{abstract}
We introduce a new technique for generating retinal fundus images that have anatomically accurate vascular structures, using diffusion models. We generate artery/vein masks to create the vascular structure, which we then condition to produce retinal fundus images. 
The proposed method can generate high-quality images with more realistic vascular structures and can create a diverse range of images based on the strengths of the diffusion model. 
We present quantitative evaluations that demonstrate the performance improvement using our method for data augmentation on vessel segmentation and artery/vein classification. 
We also present Turing test results by clinical experts, showing that our generated images are difficult to distinguish with real images.
We believe that our method can be applied to construct stand-alone datasets that are irrelevant of patient privacy.
\end{abstract}


\section{Introduction}
\label{sec:introduction}

\begin{figure*}[t]
    \centering
    \includegraphics[width = 0.95\linewidth]{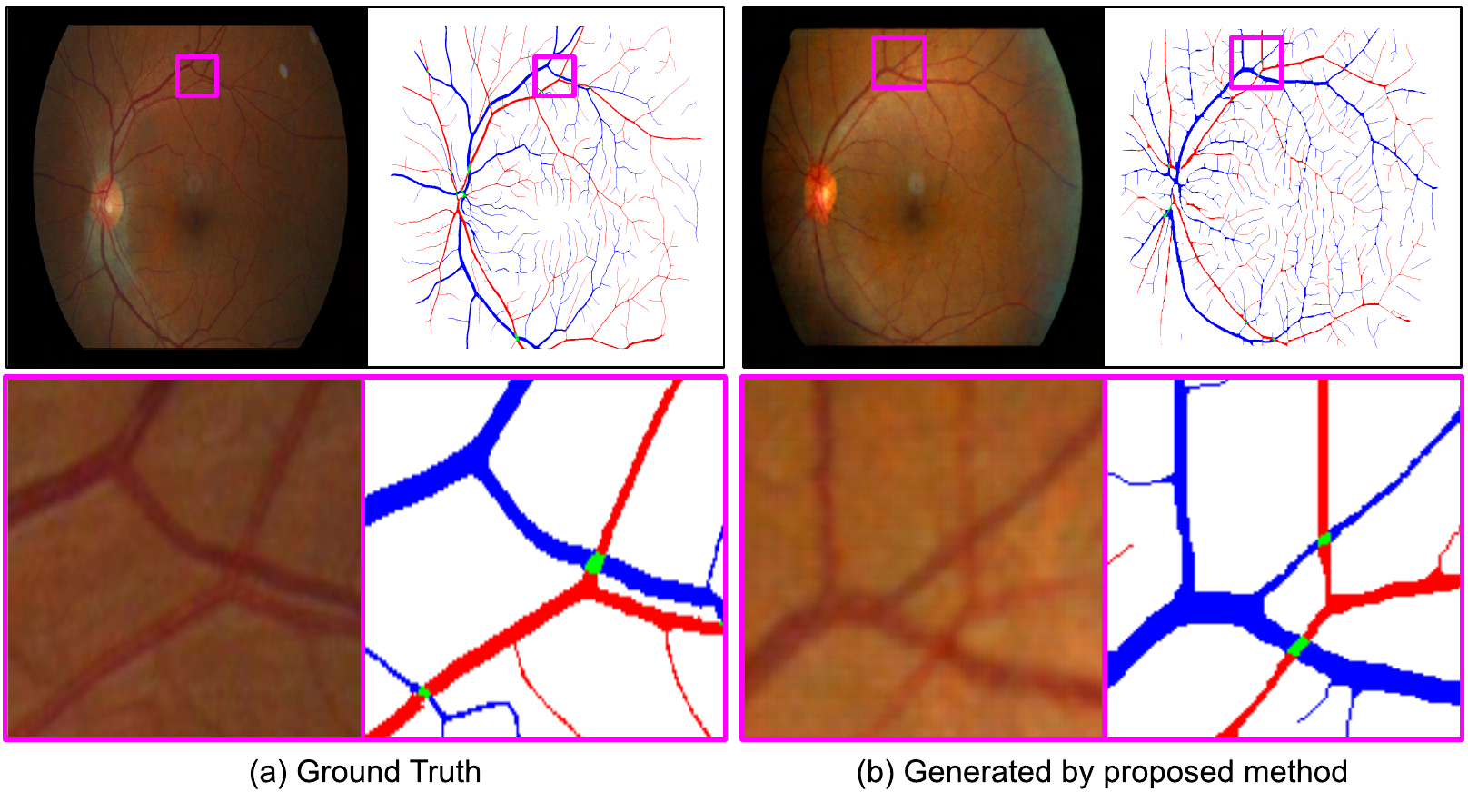}
    \caption{Sample of generated fundus image and artery/vein mask from the proposed method, together with a real data for comparison. The proposed method can generate anatomically accurate fundus images and corresponding artery/vein masks.}
    \label{fig:teaser}
\end{figure*}

\IEEEPARstart{R}etinal fundus images provide clear and high-resolution images of blood vessels in the eye. 
They are noninvasive and inexpensive. 
They are used for retinal disease diagnoses and in the early detection and prevention of chronic diseases like diabetes and hypertension.

Extensive research has been conducted on the application of machine learning to fundus images, including tasks such as vessel segmentation~\cite{Chen2021Access,Sule2022Access}, artery/vein (A/V) classification~\cite{Mookiah2021MedIA}, and disease classification~\cite{Li2021MedIA}. 
The 2D nature of these images, low noise levels, and the seemingly basic structures of the regions of interest, compared to other modalities, may have made them an attractive area of investigation. 

Despite the significant interest, limited availability of training data remains a challenge due to the difficulty in constructing the GT. 
This is particularly true for tasks such as vessel segmentation and A/V classification, which require annotating the vascular structure. 
Manual or semi-automatic annotation by clinical experts is labor-intensive, making it prohibitively expensive for large-scale datasets. 
Consequently, most publicly available datasets, including STARE~\cite{STARE}, DRIVE~\cite{DRIVE}, CHASEDB1~\cite{CHASE}, HRF~\cite{HRF} for vessel segmentation, RITE~\cite{RITE}, IOSTAR~\cite{IOSTAR} for A/V classification, and RETA~\cite{RETA} for both tasks, comprise less than one hundred images. 
Although the recently presented FIVES~\cite{FIVES} dataset comprises 800 images, the size is limited compared to datasets such as ImageNet~\cite{ImageNet} or COCO~\cite{COCO}.

Recently, a framework for vessel segmentation and A/V classification based on aligned corresponding fundus and FA images~\cite{FunFAReg_MICCAI,FunFAReg_Access,HVGN_MICCAI,HVGN_Access} was proposed. 
The results can serve as an initial baseline, but manual inspection and editing are still required before they can be deemed accurate GT vessel annotations, thereby limiting its scalability. 
Furthermore, privacy concerns may also limit the dataset size.

To address the need for more training images, generative methods can be used to synthesize additional data. A number of methods have been proposed for this purpose using various GAN models~\cite{pix2pix,PGGAN}. Early methods, such as those by Zhao et al.\cite{Zhao2018MedIA} and Costa et al.\cite{Costa2018TMI}, involved a two-step synthesis process comprising vessel mask generation and fundus image generetation. 
The first step was to synthesize the vessel segmentation mask, using adversarial autoencoders, followed by color fundus image generation using style transfer or conditional GAN. 
More recent methods, such as those by Yu et al.\cite{Yu2019Biomed} and Paolo et al.\cite{Paolo2022Elec}, have built on these earlier works, refining the models used to generate the vessel mask or fundus image. 

Kim et al.\cite{Kim2022SciRep} proposed a one-step approach, training a StyleGAN\cite{StyleGAN} on a large dataset of fundus images, that generates retinal images.
While their approach generated high-quality images, it lacks the ability to generate GT vessel masks. 

Recently, diffusion models~\cite{Dickstein2015PMLR,Ho2020NeurIPS} have been gaining much attention as an effective generative methodology and an alternative to GANs. Diffusion models often provide better stability in training compared to GANs~\cite{Dhariwal2021NeurIPS}.

Thus, we propose a novel method for generating retinal fundus images with anatomically realistic vascular structures using diffusion models. Our approach consists of two main steps: (1) unconditional generation of A/V masks to establish the vascular structure, (2) generation of retinal fundus images conditioned on the A/V mask with resolution enhancement of the generated image. To address the sparsity of the A/V mask in the first step, we introduce a modified loss function for the diffusion model. Sample results of our method are depicted in Fig.~\ref{fig:teaser}.

The main contributions of the proposed method are as follows:
\begin{itemize}
    \item The proposed method can generate fundus images with more realistic vessel structures since it is trained to generate A/V masks, rather than vessel masks. 
    \item It also generates a wider variety of high-quality images by leveraging the strengths of the diffusion model.
    \item The generated fundus image can be used directly as GT for vessel segmentation and A/V classification by assigning the generated A/V mask as the GT.
\end{itemize}

\section{Methods}\label{sec:methods}

\begin{figure*}[t!]
    \centering
    \includegraphics[width = \linewidth]{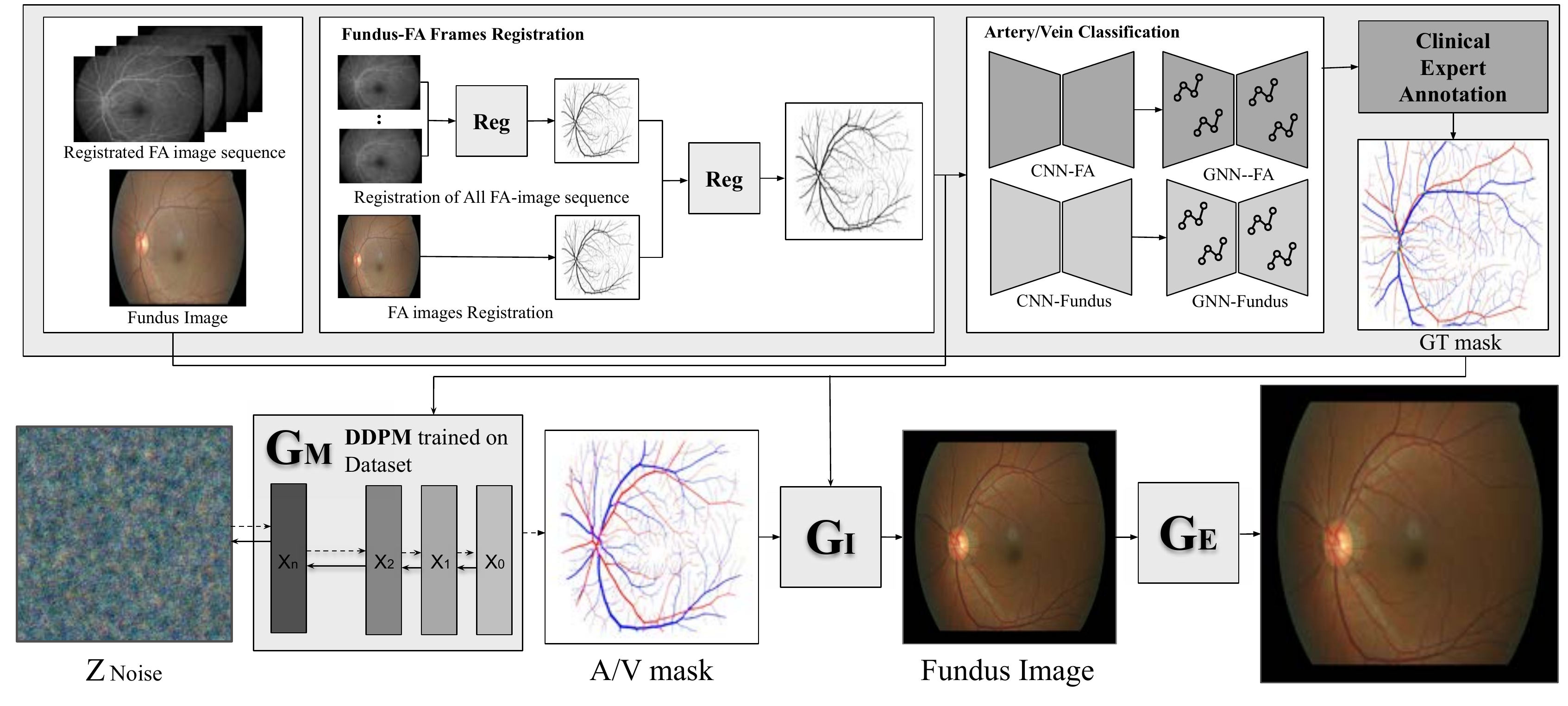}
    \caption{Overview of the proposed method. The top row depicts the process for generating GT A/V masks, which are used to train $G_{M}$, which generates new A/V masks, and $G_{I}$ which generates fundus images conditioned on the A/V masks. $G_{E}$ performs super-resolution to generate high-resolution images.}  
    \label{fig:overview}
\end{figure*}
    
In this section, we provide a detailed description of our proposed method, including the steps involved in collecting the training data. We present a graphical overview of the proposed approach in Figure~\ref{fig:overview}. 

\subsection{Constructing the training dataset}\label{sec:GT_const}

To implement the proposed method, it is necessary to have access to training datasets consisting of both fundus images and corresponding A/V masks. To generate high-quality A/V masks, we adopt the approach developed by Noh et al.~\cite{FunFAReg_MICCAI,FunFAReg_Access,HVGN_MICCAI,HVGN_Access}, which utilizes fundus images and fluorescein angiography (FA). In the following, we provide a brief overview of this process, which can be divided into two sub-processes for constructing the GT vessel mask~\cite{FunFAReg_MICCAI,FunFAReg_Access} and A/V masks~\cite{HVGN_MICCAI,HVGN_Access}.

To construct the GT vessel mask, we first construct an aggregated vessel mask from the FA sequence frames. 

Rigid registration is performed to register all frames in the FA image sequence into a common reference frame. 

Then, vessel segmentation is performed for all frames using a pre-trained CNN.
The per-frame alignments are fine-tuned using non-rigid registration on the vessel masks. The aggregate mask is constructed as the maximum vessel probability for each pixel.

The aggregated FA vessel mask is then aligned with the corresponding fundus image using another registration process, and then subjected to simple post-processing for binarization. 
Because the FA reveals fine vessels more clearly than the fundus images, this mask contains more fine details than masks generated from fundus images alone. 
Finally, this baseline vessel mask is manually edited by clinical experts to remove any errors, resulting in the GT vessel mask.

The A/V mask is constructed using the hierarchical vessel graph network (HVGN), which comprises a series of a CNN and a GNN. 
First, CNN features are constructed for the fundus and FA sequence. 
The GT vessel mask is used to extract vessel pixels, which are assigned as the vertices of a GNN, with the attributes defined as the CNN features. Edges are assigned between pixels within local window neighborhoods. 
The CNN and GNN are trained end-to-end with GT A/V masks, so that each node is classified as a label of either artery or vein. 
We note that each A/V mask is a two-channel binary mask with each channel for the artery mask and vein mask, respectively.
This baseline A/V mask is again manually edited by clinical experts to remove any errors, resulting in the final GT A/V mask.

\subsection{Generating the vascular structure}

We train the model to generate vascular structure, denoted as $G_{M}$, on the set of the GT A/V masks. We use the denoising diffusion probabilistic model (DDPM) by \cite{Ho2020NeurIPS} and improved by \cite{Dhariwal2021NeurIPS}. 

The concept of diffusion was inspired by the diffusion process in thermodynamics~\cite{Dickstein2015PMLR}. Given an image ${\bf x}_{0}$, we gradually diffuse its structure by iteratively adding random Gaussian noise. It can be shown that this forward diffusion process can be converted into a single equation to generate the diffused image at any timestep t:
\begin{equation}
    x_{t} = \sqrt{\bar{\alpha}_{t}} {\bf x_{0}} + \sqrt{1-\bar{\alpha}_{t}} {\bf \epsilon},~~~~\textrm{with}~{\bf \epsilon} \sim \mathcal{N}(0,I),
\end{equation}
where $\bar{\alpha}_{t} \coloneqq \prod_{s=1}^{t} {\alpha_{s}}$ and $\alpha_{t} \coloneqq 1-\beta_{t}$, where $\beta_{t}$ is the variance of the noise at each timestep $t$. 

The idea is to learn the process to reverse this iterative diffusion, i.e., the reverse denoising diffusion process, by training a neural network with parameters $\theta$ that generates an estimate of the noise ${\bf \epsilon}_{\theta}$ for a given image ${\bf x}_{t}$. In \cite{Ho2020NeurIPS}, a U-Net~\cite{UNet} modeled is used, and is trained with the loss function:
\begin{equation}
    L_{simple}(\theta) = \mathbb{E}_{t,{\bf x_{0}},{\bf \epsilon}}\left[ \lVert {\bf \epsilon} - {\bf \epsilon}_{\theta} \left( \sqrt{\bar{\alpha}_{t}} {\bf x_{0}} + \sqrt{1-\bar{\alpha}_{t}} {\bf \epsilon}, t \right) \rVert^{2} \right].
\end{equation}

In our experiments, we found that if we directly train $G_{M}$ on the A/V masks using the above configuration, many generated samples contained little to no vascular structures at all. 
As A/V masks are mostly composed of background, with artery and veins comprising only a small portion of the whole area, even an empty image can be a reasonable sample if all pixels are considered equally. We thus modifiy the loss function $\mathcal L$ as follows:
\begin{equation}
    \footnotesize
    L_{vessel}(\theta) = \mathbb{E}_{t,{\bf x_{0}},{\bf \epsilon}}\left[ \exp{(c{\bf x_{0}})} \odot \lVert {\bf \epsilon} - {\bf \epsilon}_{\theta} \left( \sqrt{\bar{\alpha}_{t}} {\bf x_{0}} + \sqrt{1-\bar{\alpha}_{t}} {\bf \epsilon}, t \right) \rVert^{2} \right],
\end{equation}
where ${\bf x_{0}}$ is the given A/V mask, and $\exp{({\bf x_{0}})}$ is the pixelwise exponential of each pixel value, $c$ is a scalar parameter, and $\odot$ denotes the Hadamard, or element-wise, product. Also, we empirically determined the resolution of the model output and thereby the generated images as $256 \times 256$, as the model was difficult to train for higher resolutions. Note that, we incorporate the increased number of attention heads and attention layers, and the adaptive group norm layers as proposed in \cite{Dhariwal2021NeurIPS}.

\subsection{Generating the fundus image}

Given the A/V masks, we then train the model to generate the fundus image, denoted as $G_{I}$. We apply a conditional GAN, namely, the pix2pixHD~\cite{pix2pixHD} as the network model. It is an improvement over the previous pix2pix~\cite{pix2pix} by using a coarse-to-fine generator, a multi-scale discriminator architecture, and a robust adversarial learning objective function. $G_{I}$ is trained with GT A/V masks as the input training set and the corresponding fundus images as the output supervision. Since generated A/V masks with low-resolution $256 \times 256$ will be given as input at inference time, we resize the GT A/V masks and fundus images to have the same size.

To enhance the low-resolution of the generated images, we further train a super-resolution network, denoted as $G_{E}$, to up-sample the images at rate 4 to $1024 \times 1024$. We apply the Enhanced Deep Residual Network (EDSR)~\cite{EDSR}. The GT fundus images are used as supervision for corresponding input images down-sampled by rate 4.

\section{Experiments}\label{sec:exp}

\subsection{Dataset and implementation details}

\noindent
{\bf Private Dataset:} We use a private dataset of corresponding fundus images and FA comprising 449 cases of corresponding fundus image and FA pairs from 339 patients from Seoul National University Bundang Hospital (SNUBH)\footnote {This study was approved by the Institutional Review Board (IRB) of the Seoul National University Bundang Hospital (IRB no. B-1810-501-104, approval date 12 Oct, 2018), and the requirement of informed consent was waived from the IRB. The study complied with the guidelines of the Declaration of Helsinki.}. Canon CF60Uvi, Kowa VX-10, and Kowa VX-10a cameras were used to acquire the fundus images and FA. Image resolutions were all normalized to $1536 \times 1024$ from original resolutions, varying from $1604 \times 1216$ to $2144 \times 1424$. We note that we have named this dataset as the FIREFLY (Fundus Images REgistered with FLuorescein angiographY) dataset, and will refer to it as such.

For training the CNN for vessel segmentation within the process for constructing the GT vessel segmentation masks of our private SNUBH dataset, we use the public datasets DRIVE~\cite{DRIVE} and HRF~\cite{HRF}. For training the HVGN for constructing the GT A/V masks, we use 46 cases among the 449 of the private dataset. 

\noindent
{\bf Public Dataset:} We also trained and evaluated our method on the DRIVE public dataset, which includes expert annotated vessel mask GT. Annotated artery/vein vessel mask GT were also made available from added research, entitled as the RITE dataset \cite{RITE}. 

The dataset consists of 20 images each for the training and testing datasets.

\noindent
{\bf Environment and implementation details:} Experiments were performed on a system with a 2.2 GHz Intel Xeon CPU and 4 nVidia Titan V GPUs. We used the DDPM implementation publicly provided by \cite{Dhariwal2021NeurIPS}. We set the maximum number of timestep iterations $T = 1000$ for sampling the DDPM. 

\subsection{Qualitative Evaluations}

\begin{figure*}[hbtp]
    \centering
    \includegraphics[width = 0.8\linewidth]{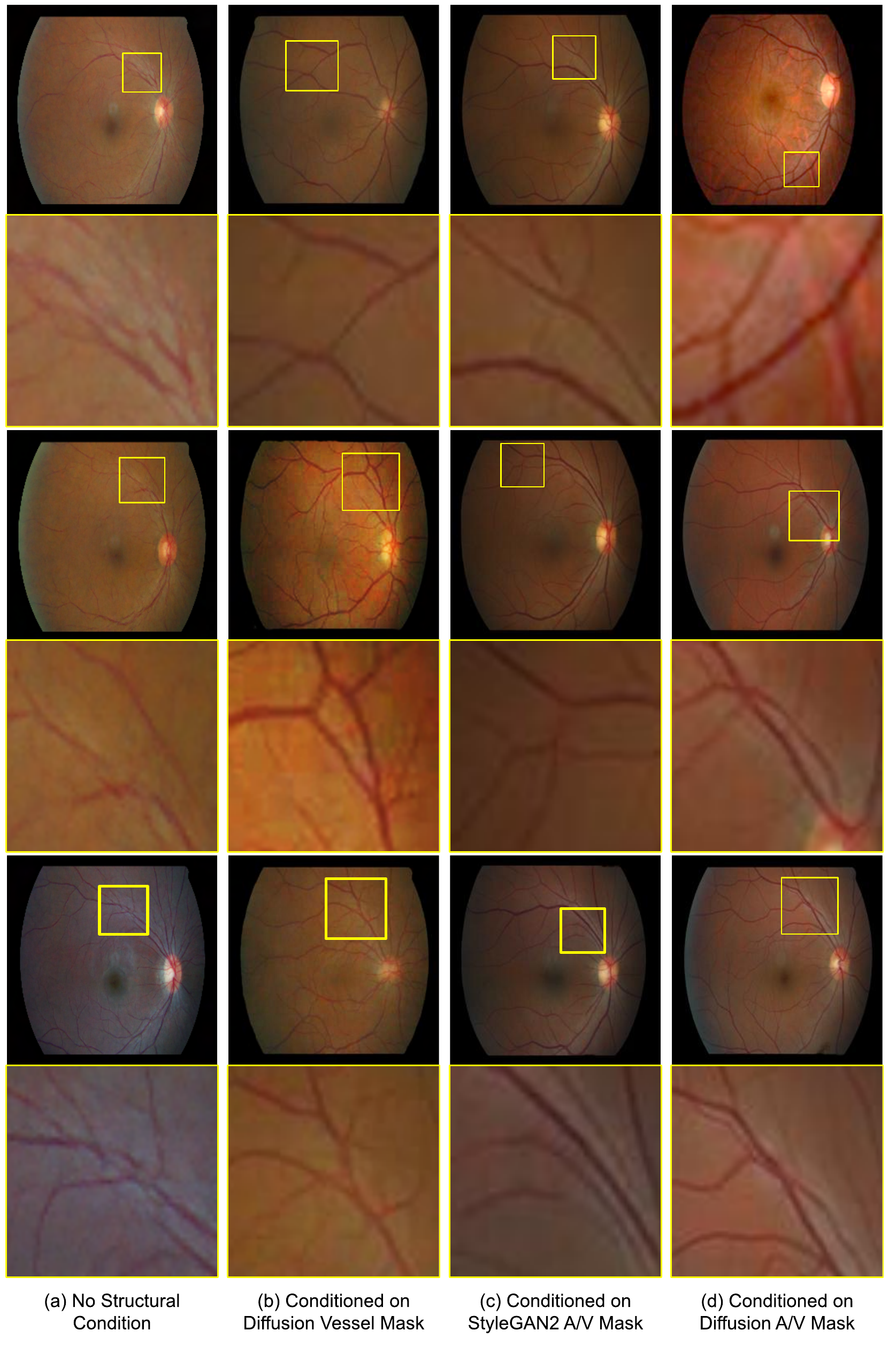}
    \caption{Samples of generated fundus images for qualitative evaluation. Each column depicts generated fundus images (1) conditioned on A/V masks generated by StyleGAN2, (2) from diffusion model directly trained on fundus images, (3) conditioned on vessel masks generated by diffusion model, (4-proposed) conditioned on A/V mask generated by diffusion model.}  
    \label{fig:qual_panel}
\end{figure*}

\noindent
{\bf Qualitative visual comparisons:}
We first present qualitative examples of the generated images in Fig.~\ref{fig:qual_panel}. Each column presents three samples generated using a diffusion model, (a) without any structural conditions, (b) conditioned on A/V masks from StyleGAN2, (c) conditioned on vessel masks from a diffusion model, and (d) the proposed method of images conditioned on A/V masks from a diffusion model. For each sample, we present the global image (best viewed using a monitor with by zoom-in) above and a crop-and-zoomed patch below. The zoomed patches for columns (a), (b), and (c) all depict vascular structures with anatomically infeasible structures, including fragmented vessels, unrealistic bifurcations or trifurcations, or unnatural loops. In general, we observed that the generated images from the proposed method generally has more realistic vascular structures.

\begin{figure*}[t]
    \centering
    \includegraphics[width = 1\linewidth]{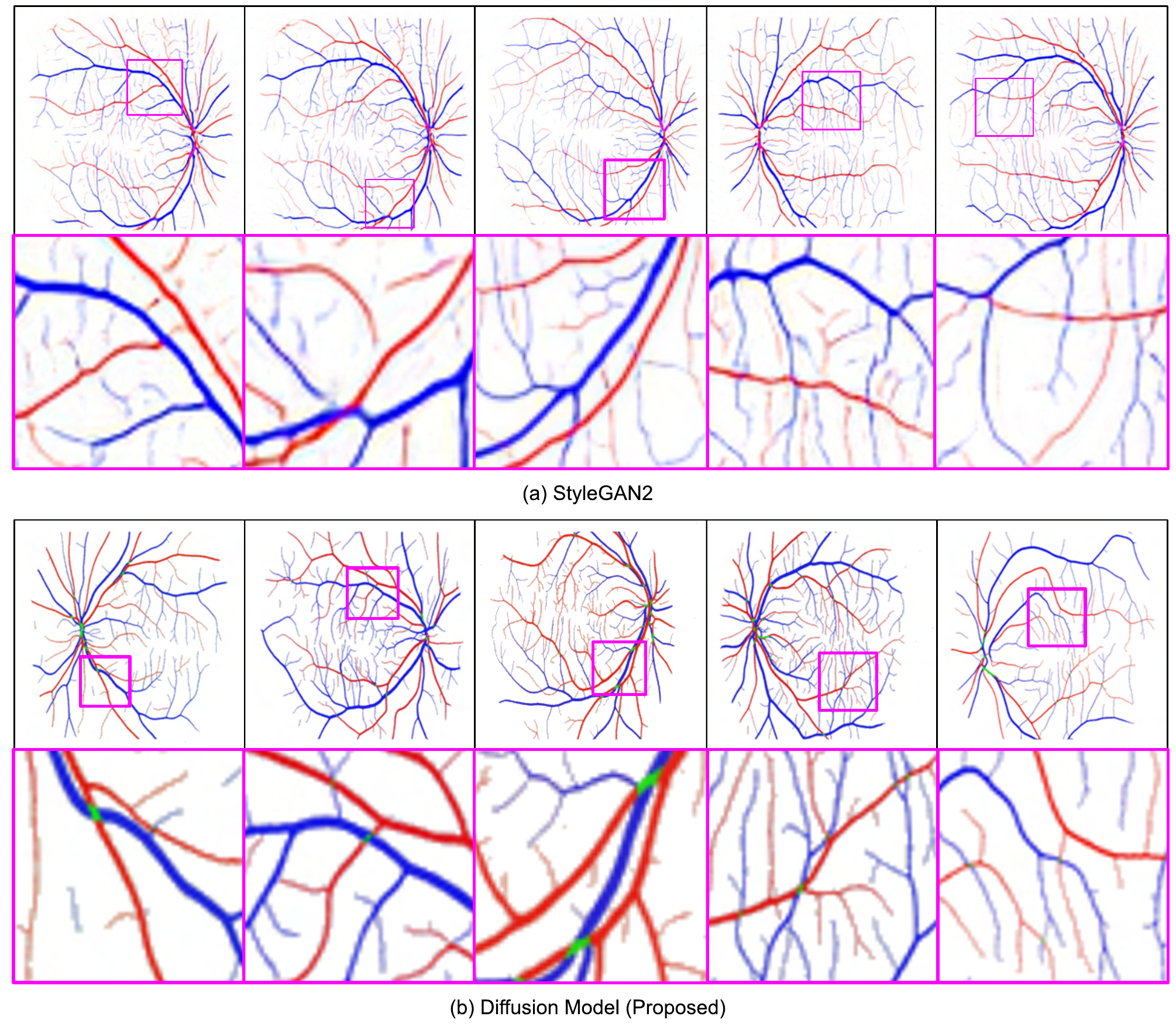}
    \captionsetup{justification=centering}
    \caption{Samples of generated artery/vein masks from StyleGAN2 and the diffusion model.}  
    \label{fig:av_masks}
\end{figure*}

We also present comparisons between the A/V masks generated from (a) StyleGAN2 and (b) our trained diffusion model in Fig.~\ref{fig:av_masks}. We can observe that StyleGAN2 mostly fails in generating contiguous vascular structures, especially in the vessel branching, and cannot model crossings. In contrast, A/V masks generated by diffusion have realistic bifurcations and crossings. Also, results from StyleGAN2 are generally more noisy and cannot be readily binarized, compared to images generated by the diffusion. 

\noindent
{\bf Qualitative Turing test:}
We next present results of qualitative Turing tests. We asked five board certified ophthalmologists to determine if the image is from a real patient or was synthesized for (1) real fundus images acquired from patients, (2) fundus images generated from a diffusion model directly trained on fundus images, (3) fundus images generated from A/V masks generated using StyleGAN2~\cite{StyleGAN2}, and (4) fundus images generated from A/V masks generated using a diffusion model. For each category, 20 images were presented. 

Table~\ref{tab:turing} summarizes the results. It is evident that there are no decisive image features for identifying real fundus images, as the experts mistook real images as synthetic images in 40\% of the cases. Nonetheless, they were able to correctly identify 90\% of the images generated from diffusion without generating vascular structure, likely due to unrealistic artifacts. For images generated from the proposed method, the participants were incorrect in more than half the cases. For images generated using StyleGAN2~\cite{StyleGAN2}, the expert participants guessed that a higher portion were real compared to the real fundus images. We believe that these results demonstrate that more realistic fundus images are generated when conditioned on vascular structures.

\begin{table*}[t]

\caption{Qualitative comparative evaluation of generated fundus images via Turing test, asking whether the image is real or synthesized. Each test comprises 20 images.}
\label{tab:turing}
\centering
\begin{tabular}{c|c|c|c}
    \hline
    Real/Synth. & Gen. Model & Vascular structure & Acc.(95\% CI) \\
    \hline
    Real & - & - & 60($\pm 11.47$) \\\hline
    Synth & Diffusion~\cite{Dhariwal2021NeurIPS} & None & 90($\pm 8.99$) \\\hline
    Synth & Diffusion~\cite{Dhariwal2021NeurIPS} & A/V mask & 41.25($\pm 7.35$) \\\hline
    Synth & StyleGAN2~\cite{StyleGAN2} & A/V mask & 22.5($\pm 9.34$) \\\hline
    \bottomrule
\end{tabular}
\end{table*}

\subsection{Quantitative Evaluations}\label{sec:public_data}

As it was observed in the qualitative Turing test, it is very difficult to estimate or measure how realistic is the generated image. We thus focus on measuring the effect of the generated images on the original goal, of improving the training of methods of vessel segmentation and A/V classification.

In Table~\ref{tab:quant_comp}, we present the performance of recent state-of-the-art methods SSA-Net~\cite{SSANet} and RV-GAN~\cite{RV_GAN} for vessel segmentation, and HVGN~\cite{HVGN_Access} for A/V classification, with and without data augmentation based on the generated fundus images and corresponding vessel masks. For all methods, we use the implementations provided by the authors. For the HVGN method, we evaluate the version with only fundus image inputs, in contrast to the version that with both fundus and FA images, which was used to construct the GT A/V masks of the FIREFLY dataset. 

We present results using the private FIREFLY dataset and the public DRIVE~\cite{DRIVE}/RITE~\cite{RITE} dataset. 
For the FIREFLY dataset, we assign a random partition of 100 images for the test set and the remaining 359 images for the training set. 
We present results for within-domain evaluation, i.e., using the training and test sets from the same dataset, as well as the cross-domain evaluation trained on the FIREFLY training set and tested on the DRIVE/RITE test set.
For all configurations, 3000 synthetic A/V masks and images and were generated from the proposed method were used in the data augmentation.
Accuracy (Acc), sensitivity (Se), specificity (Sp), and the area-under-curve (AUC) of the receiver-operating-characteristic (ROC) curve are measured based on pixel-wise true/false positives and true/false negatives, for binary vessel masks and A/V masks comprising binary artery and vein tree masks, respectively.

We observed that the quantitative results were improved for all measurements when using the generated data augmentation, regardless of the vessel analysis task or specific method, or the specific train/test dataset configuration. We believe these results strongly support the effectiveness and generalizability of the proposed method. 

In Table~\ref{tab:segm_cls_tr_size}, we present the comparative evaluation according to the number of the synthesized images for data augmentation, for the FIREFLY dataset. 
Similar to a practical dataset expansion process, we simulate a gradual increase of the dataset, adding 1000 generated images up to a maximum of 3000 images. 
We train methods for vessel segmentation and A/V classification, SSA-Net~\cite{SSANet} and HVGN~\cite{HVGN_Access}, respectively, on these datasets.
We observed that the quantitative results were improved for almost all measurements with the increase in the synthetic data.
Again, we believe that these results support the relevance of the generated images in training methods for vessel analysis.

To further evaluate how much the generated images can faithfully constitute relevant information, we trained the vessel segmentation and A/V classification methods on synthetic images only, without any real images.
We observed that there were only slight decreases in the quantitative results, demonstrating that the generated images faithfully represent the characteristics of the original real images, and contain sufficient information for training methods for vessel analysis.

\begin{figure*}[p]
    \centering
    \includegraphics[width = 1\linewidth]{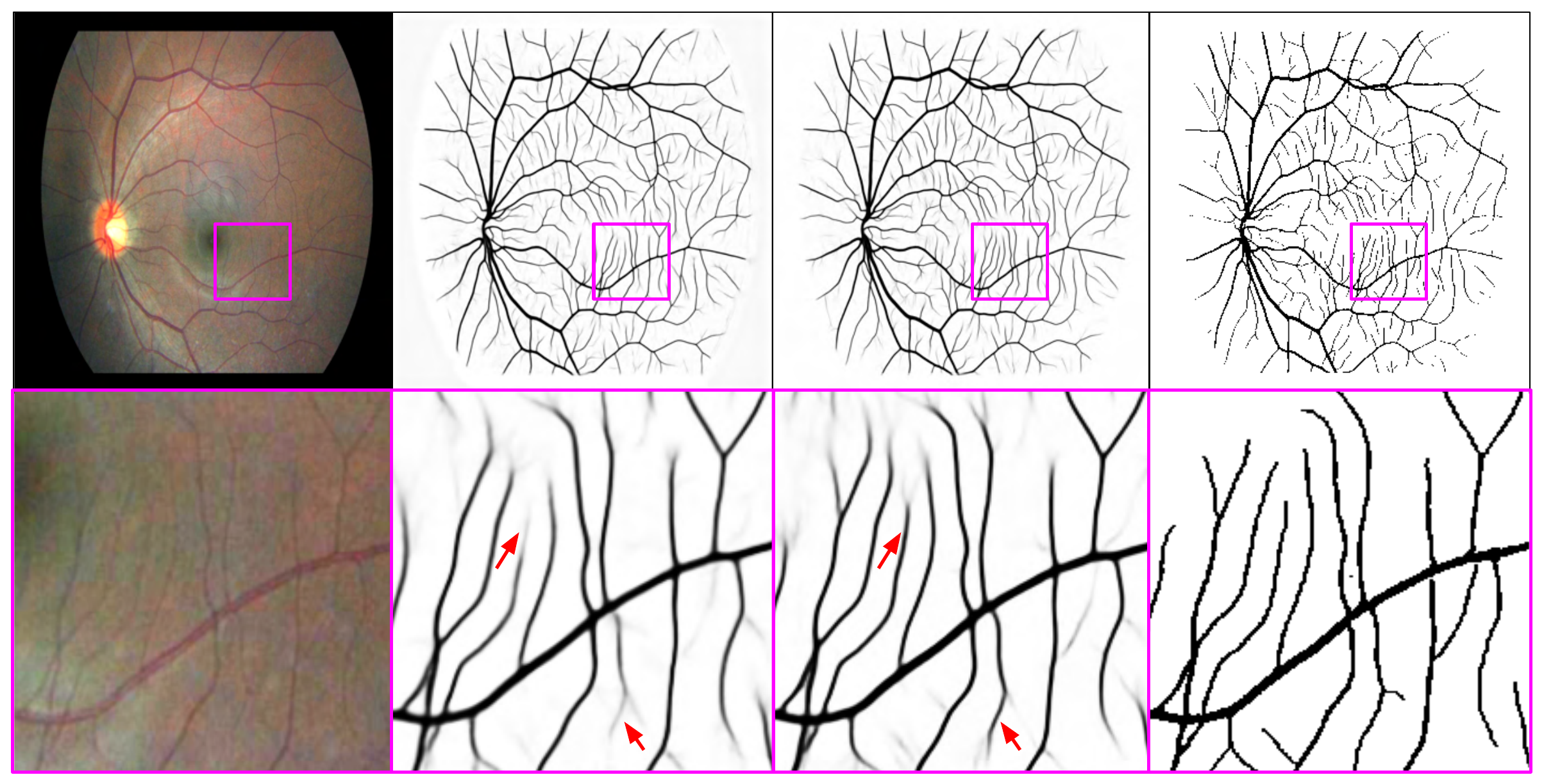}
    \caption{Sample vessel segmentation result of SSA-Net~\cite{SSANet}, trained on the FIREFLY dataset. (a) Input fundus image, (b) result of SSA-Net trained only on real images, (c) result of SSA-Net trained with data augmentation from synthetic dataset of 3000 images generated by the proposed method, and (d) the GT. The top and bottom rows show the images in full and zoomed resolution.}  
    \label{fig:result_segm}
\end{figure*}

\begin{figure*}[bh]
    \centering
    \includegraphics[width = 1\linewidth]{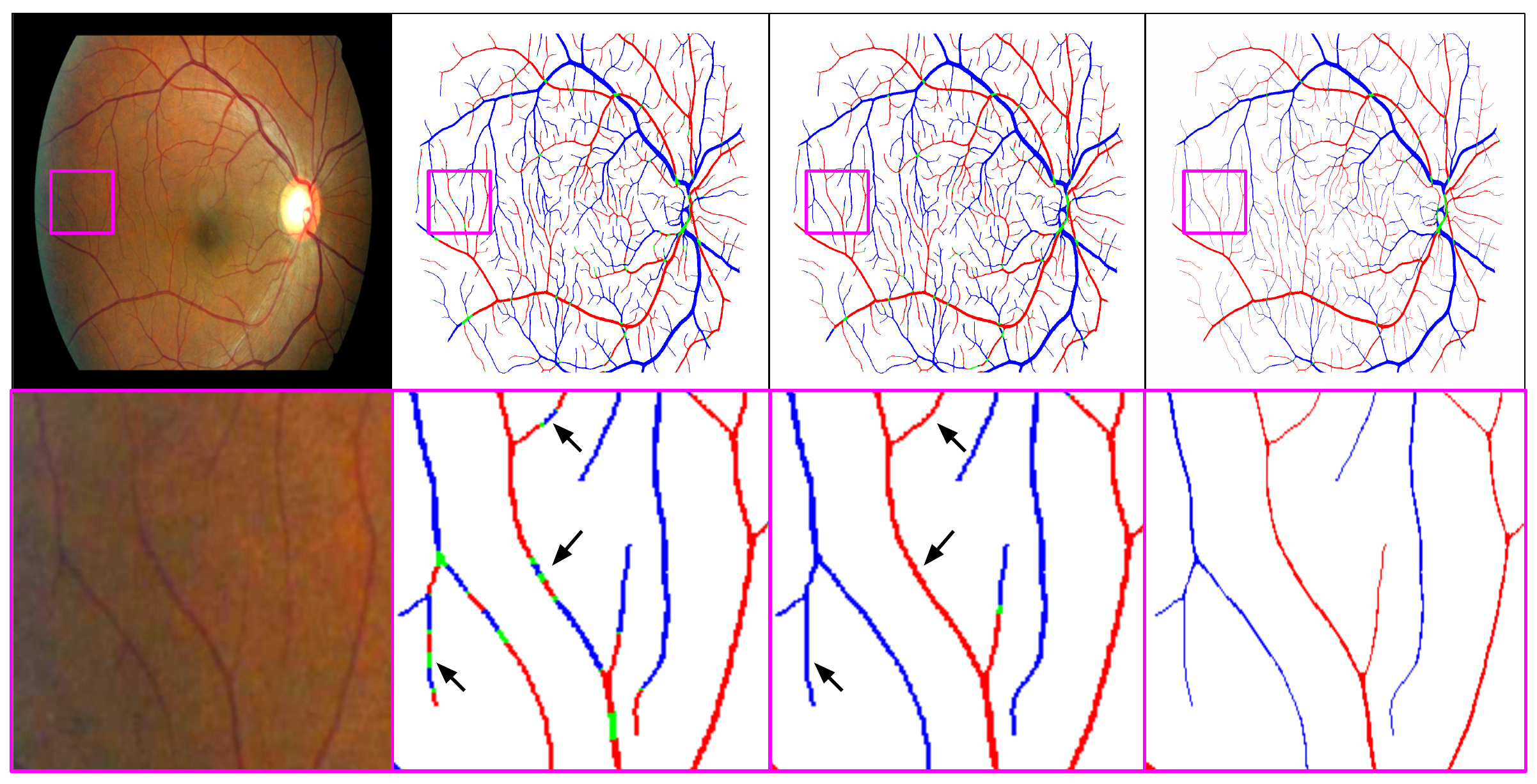}
    \caption{Sample A/V classification result of HVGN~\cite{HVGN_Access}, trained on the FIREFLY dataset. (a) Input fundus image, (b) result of HVGN trained only on real images, (c) result of HVGN trained with data augmentation from synthetic dataset of 3000 images generated by the proposed method, and (d) the GT. The top and bottom rows show the images in full and zoomed resolution.}  
    \label{fig:result_av}
\end{figure*}

We provide sample results for vessel segmentation and A/V classification in Fig.~\ref{fig:result_segm} and Fig.~\ref{fig:result_av}, respectively.
These results demonstrate that the data augmentation from the images generated from the diffusion model helps to correct many misclassifications and broken vessel branches in vessel segmentation and A/V classification.

\begin{table*}[t]
\caption{Comparative evaluation of vessel analysis methods trained w/wo generative data augmentation using on the proposed method. Area-under-curve (AUC) of the receiver-operation-characteristic curve (ROC), Accuracy (Acc), Sensitivity (Se), and specificity (Sp) are presented.}
\label{tab:quant_comp}
\centering
{
\begin{tabular}{c|c|c|c|c|cccc}
    \toprule
    Task & Model & Train DB$^*$ & Aug.& Test DB & AUC & Acc. & Se. & Sp. \\
    \hline
     \multirow{10}{*}{Segmentation} & \multirow{6}{*}{SSA-Net~\cite{SSANet}} & \multirow{2}{*}{FIREFLY} & \xmark & \multirow{2}{*}{FIREFLY} & 0.973 & 0.963 & 0.797 & 0.983  \\
      & & &  \checkmark & & 0.984{\scriptsize{\textcolor{red}{+.011}}} & 0.976 & 0.815 & 0.990 \\
      \cline{3-9}
    & & \multirow{2}{*}{FIREFLY} & \xmark & \multirow{2}{*}{DRIVE-AV} & 0.948 & 0.910 & 0.870 & 0.920 \\
     & & & \checkmark & & 0.952{\scriptsize{\textcolor{red}{+.004}}} & 0.918 & 0.874 & 0.924 \\
      \cline{3-9}
     & & \multirow{2}{*}{DRIVE-AV} & \xmark & \multirow{2}{*}{DRIVE-AV} &  0.961 & 0.914 & 0.868 & 0.921 \\
      & & & \checkmark & & 0.966{\scriptsize{\textcolor{red}{+.005}}} & 0.921 & 0.906 & 0.923 \\
      \cline{2-9}
     & \multirow{4}{*}{RV-GAN~\cite{RV_GAN}} & \multirow{2}{*}{FIREFLY} & \xmark & \multirow{2}{*}{FIREFLY} & 0.937 & 0.945 & 0.713 & 0.989 \\
      & & & \checkmark & & 0.944{\scriptsize{\textcolor{red}{+.007}}} & 0.947 & 0.672 & 0.993 \\
      \cline{3-9}
     & & \multirow{2}{*}{DRIVE-AV} & \xmark & \multirow{2}{*}{DRIVE-AV} & 0.970 & 0.950 & 0.630 & 0.919 \\
     & & & \checkmark & & 0.981{\scriptsize{\textcolor{red}{+.011}}} & 0.967 & 0.658 & 0.995 \\
      \hline
     \multirow{6}{*}{A/V cls.} &\multirow{6}{*}{HVGN~\cite{HVGN_Access}} & \multirow{2}{*}{FIREFLY} & \xmark & \multirow{2}{*}{FIREFLY} & 0.973 & 0.912 & 0.923 & 0.901\\
     & & & \checkmark & & 0.980{\scriptsize{\textcolor{red}{+.007}}} & 0.922 & 0.930 & 0.915 \\
     \cline{3-9}
     & & \multirow{2}{*}{FIREFLY} & \xmark & \multirow{2}{*}{DRIVE-AV}  & 0.909 & 0.819 & 0.971 & 0.768 \\
     & & & \checkmark & & 0.914{\scriptsize{\textcolor{red}{+.005}}} &  0.824 & 0.879 & 0.769\\
       \cline{3-9}
     & & \multirow{2}{*}{DRIVE-AV} & \xmark & \multirow{2}{*}{DRIVE-AV} & 0.917 & 0.855 & 0.877 & 0.832 \\
      & & & \checkmark & & 0.924{\scriptsize{\textcolor{red}{+.007}}} & 0.860 & 0.884 & 0.835\\
    \bottomrule
    \multicolumn{9}{p{370pt}}{*{\scriptsize{The diffusion model for image generation and the model for vessel analysis tasks are both trained using the same \emph{Train DB}. A total of 3000 synthetic A/V masks and fundus images generated by the proposed method were used for data augmentation.}}}
\end{tabular}
}
\end{table*}

\begin{table*}[t]
\caption{Comparative evaluation of vessel analysis methods on the FIREFLY dataset, trained with different number of generated images in data augmentation. Area-under-curve (AUC) of the receiver-operation-characteristic curve (ROC), Accuracy (Acc), Sensitivity (Se), and specificity (Sp) are presented.}
\label{tab:segm_cls_tr_size}
\centering
    \begin{tabular}{c|cccc|cccc}
    \toprule
    \multirow{3}{*}{Training Set Size } & \multicolumn{4}{c|}{Segmentation} & \multicolumn{4}{c}{A/V cls.} \\
     & \multicolumn{4}{c|}{SSA-Net~\cite{SSANet}} & \multicolumn{4}{c}{HVGN~\cite{HVGN_Access}} \\
    \cline{2-9}
          &AUC & Acc. & Se. & Sp. & AUC & Acc. & Se. & Sp. \\ 
    \hline
     Real. 359 & 0.973 & 0.963 & 0.797 & 0.983 & 0.973 & 0.912 & 0.923 & 0.901 \\
     Real + Synth. 500 & 0.979 & 0.972 & 0.798 & 0.986 & 0.976 & 0.915 & 0.926 & 0.905 \\
     Real + Synth. 1000 &  0.980 & 0.973 & 0.803 & 0.988 & 0.976 & 0.915 & 0.927 & 0.902 \\
     Real + Synth. 2000 & 0.983 & 0.975 & 0.812 & 0.989 & 0.978 & 0.921 & 0.928 & 0.914 \\
     Real + Synth. 3000 & \bf0.984 & \bf0.976 & 0.815 & \bf0.990 & \bf0.980 & \bf0.922 & \bf0.930 & \bf0.915 \\
     Synth. 3000 & \bf0.984 & 0.975 & \bf0.816 & 0.988 & \bf0.980 & 0.921 & 0.929 & 0.912 \\
     Synth. 2000 & 0.982 & 0.974 & 0.811 & 0.989 & 0.978 & 0.919 & 0.926 & 0.910 \\
     Synth. 1000 &  0.979 & 0.973 & 0.801 & 0.988 & 0.975 & 0.914 & 0.924 & 0.903 \\
        \bottomrule
\end{tabular}
\end{table*}

\noindent
{\bf Quantitative evaluation of perceptual image quality:}
In Table~\ref{tab:FID}, we present the FID values for the fundus images generated from A/V masks generated using StyleGAN2~\cite{StyleGAN2}, and those generated from A/V masks generated using diffusion~\cite{Dhariwal2021NeurIPS}. 
It is evident that the diffusion model results in lower distances, demonstrating the benefit of the diffusion model.
We note that, due to computational constraints, neither GAN and diffusion model were not optimized in respect to the scores, as was done in \cite{Kim2022SciRep}.

\begin{table}[t]
\small
\caption{Fréchet Inception Distance (FID) for generated fundus images.}
\label{tab:FID}
\centering
\begin{tabular}{c|c|c}
    \hline
    Gen. Model & Vascular structure & FID \\
    \hline
    StyleGAN2~\cite{StyleGAN2} & A/V mask & 122.8 \\
    Diffusion~\cite{Dhariwal2021NeurIPS} & A/V mask & 86.78 \\
    \hline
\end{tabular}
\end{table}

\section{Discussion}\label{sec:discussion}

\noindent

Different models may be used as the generators for the mask, $G_{M}$, the mask-to-image generator $G_{I}$, and the high-res image generator $G_{E}$.
For instance, diffusion models could also be used for both $G_{I}$ and $G_{E}$. 
Here, we avoided their use, because they required considerably higher computation, while only achieving comparable results. 
This may be due to the nature of the GT fundus images, which were limited to images from normal patients, and mostly contains smooth textures other than the vascular structure.
As our main target application was the use of generated images for training methods for vessel analysis, we did not consider various pathologies such as macular degeneration.
Thus, further research may be warranted for improving the generation framework when generating synthetic datasets for disease diagnosis.

In the short term, we plan to apply the proposed method to construct a large-scale dataset with high-quality GT vascular structures and without any privacy concerns. 
Based on the dataset, we hope to further develop more advanced metrics for measuring the performance of segmentation and classification techniques.

\bibliographystyle{IEEEtran}
\bibliography{manuscript}

\end{document}